\documentclass[a4paper]{article}
\usepackage{multicol}
\usepackage{graphics}
\usepackage{amsmath}
\usepackage{psfig} 
\renewcommand{\title}[1]{\ignorespaces{\centering{\large{\bf{#1}}}\vskip1.5pc}}
\renewcommand{\author}[1]{\ignorespaces{\centering{#1}\vskip2.5pt}}
\newcommand{\address}[1]{\ignorespaces{\centering{\small{\it{#1}\vskip1.8pc}}}}

\renewcommand{\section}[1]{\ignorespaces{\vskip1.0pc{\bf{#1}}\vskip1.0pc}}
\renewcommand{\subsection}[1]{\ignorespaces{\vskip1.0pc\centering{\bf{#1}}\vskip0.3pc}}
 
\begin{document}
\parindent0mm
\title{GRAVITOMAGNETIC EFFECTS}
\author{\bf G.~Sch\"afer}

\bigskip

\address{Friedrich-Schiller-Universit\"at Jena,
Theoretisch-Physikalisches Institut\\
Max-Wien-Pl. 1, 07743 Jena, Germany, Email: gos@tpi.uni-jena.de}

\medskip

\setlength{\columnsep}{7mm}
\begin{multicols}{2}[][85mm]
{\bf ABSTRACT/RESUME}

\medskip

The paper summarizes the most important effects in Einsteinian
gravitomagnetic fields related to propagating light rays,
moving clocks and atoms, orbiting objects,
and precessing spins. Emphasis is put onto the gravitational
interaction of spinning objects. The gravitomagnetic field
lines of a rotating or spinning object are given in analytic form.

\bigskip

\section{1.~INTRODUCTION}
In the Einstein theory of gravity the gravitational field is described
by ten potential functions $g_{\mu\nu}$ ($\mu, \nu = 0,1,2,3$,
$g_{\mu\nu} = g_{\nu\mu}$) which
at the same time are the metric coefficients of curved spacetime.
The potential function $g_{00}$, sometimes called gravitational
redshift potential, is connected with the Newtonian
gravitational potential, the six functions $g_{ij}$ ($i,j = 1,2,3$) describe
the geometry of the curved three-dimensional spaces defined by the
slices of constant time $t$, or $x^0=ct$, where $c$ denotes the speed
of light, and finally, the three functions $g_{0i}$ describe how the geometry
of the three-dimensional slices rotates in going from one slice to
another. For weak gravitational fields, the three potential functions
$g_{0i}$ behave very much like the electromagnetic 3-potential
$A_i^{\rm em}$ on account of which the $g_{0i}$ are often called 
gravitomagnetic field potentials. Similarly to the generation of the
electromagnetic potential $A_i^{\rm em}$ through charge currents is
the gravitomagnetic field generated through mass currents
(momentum densities) if also by a factor of four more efficient
because of an underlying tensor theory (spin-2 field theory versus
spin-1 field theory of  electrodynamics; implying also a sign difference). 

In this paper important effects connected with {\bf weak} gravitomagnetic fields
will be discussed. The applied class of gravitomagnetic fields will
originate both from spinning and orbiting mass currents. The objects moving or
propagating in the gravitomagnetic fields will be non-spinning and spinning
objects (celestial bodies, particles, clocks, atoms, black holes, etc.)
as well as light rays. The gravitomagnetic field lines of a spinning object
will be given in analytic form. In all publications known to the author, the
graphs for the gravitomagnetic field lines are not presented in fully exact
form.

\bigskip

\section{2.~SPINS IN MINKOWSKI SPACE}

In the weak-field limit of Einstein's theory of gravity
the treatment of spinning objects is closely related to their
treatment in Minkowski space. Therefore, in this
section, properties of spinning or (rigidly) rotating objects in
Minkowski space will be discussed. The most important outcome of the present
section is  the relation between the canonical position variable of a spinning
object and the various centre-of-mass definitions.

Written in canonical variables, the total angular momentum
of a spinning object in Minkowski space takes the form 

\medskip

\begin{equation} 
{\bf J} ~ = ~ {\bf R} \times {\bf P} ~+ ~{\bf S} \, ,
\end{equation}

\bigskip

where ${\bf R}$, ${\bf P}$, and ${\bf S}$ denote the position vector,
the linear momentum, and the spin vector of the object, respectively.  ${\bf R}$
and ${\bf P}$ are canonically conjugate varibles which 
commute with the spin vector ${\bf S}$  the components of which 
fulfil the standard angular momentum commutation relations.
The Poincar\'e algebra tells us that the centre-of-mass
constant ${\bf K}$ has to take the form [1] 

\medskip 

\begin{equation} 
{\bf K} ~= ~x^0 {\bf P}~ - ~ {\bf R} P^0
~ + ~\frac{1}{P^0+Mc}~  {\bf S} \times {\bf P}\, ,
\end{equation}

\bigskip

where $M$ and $cP^0 = H$ are the rest mass and the energy of the object,
respectively. The 4-momentum reads $P^{\mu} = (P^0, {\bf P})$
($P^{\mu}P_{\mu} = - M^2c^2$) and for
the (coordinate) velocity ${\bf v}$ the relation ${\bf v} = c {\bf P}/P^0$
holds. From the Eq. 2 the centre-of-mass coordinate results in the form

\medskip

\begin{equation} 
\hat{{\bf R}} ~= ~{\bf R} ~ -~ \frac{1}{P^0+Mc} ~{\bf S} \times \frac{{\bf
P}}{P^0}\, , 
\end{equation}

\bigskip

using the standard definition   

\medskip

\begin{equation} 
{\bf K} ~= ~x^0 {\bf P} ~- ~{\hat{\bf R}} P^0 \, . 
\end{equation}

\bigskip

In terms of the antisymmetric spin-4-tensor $S^{\mu\nu}$ the Eq. 4
implies  
$S^{0i}=0$ (so-called Corinaldesi-Papapetrou spin supplementary
condition, e.g. see [2]).
This spin tensor we may call  $\hat{S}^{\mu\nu}$ with $\hat{S}^{ij} =
\epsilon^{ijk}\hat{S}_k$, where $\epsilon^{ijk}$ denotes the total
antisymmetric Levi-Civita tensor.
Then the
total angular momentum 4-tensor $J^{\mu\nu}$ takes the form   

\medskip

\begin{equation} 
J^{\mu\nu} ~=~ \hat{X}^{\mu} P^{\nu} ~-~  \hat{X}^{\nu} P^{\mu} ~+~
 \hat{S}^{\mu\nu}, \quad  \hat{S}^{0\mu} ~=~ 0\, ,
\end{equation}

\bigskip

with $\hat{X}^{\mu} = (\hat{X}^0, \hat{{\bf R}})$.

Let us call
the spin-4-tensor say, $\hat{S}^{\mu\nu}_{\rm rf}$,
if the covariant condition holds, $S^{\mu\nu}U_{\nu} =0$  (so-called Pirani
spin supplementary condition, e.g. see [2]) with $U_{\nu} = P_{\nu}/Mc$.
The index ${\rm rf}$ is chosen such as 
to indicate that the definition of the centre-of-mass is made in 
the rest frame.
Then we get

\medskip

\begin{equation} 
 J^{\mu\nu} ~=~ \hat{X}^{\mu}_{\rm rf} P^{\nu}~ -~  \hat{X}^{\nu}_{\rm rf} P^{\mu}~ +~
 \hat{S}^{\mu\nu}_{\rm rf}, \quad \hat{S}^{\mu\nu}_{\rm rf} U_{\nu} ~=~0\,,
\end{equation}

\bigskip

with
 $\hat{X}^{\mu}_{\rm rf}
 = (\hat{X}^0_{\rm rf}, \hat{{\bf R}}_{\rm rf})$ and
 $\hat{S}^{\mu\nu}_{\rm rf} =
 \epsilon^{\mu\nu\alpha\beta}U_{\alpha}\hat{S}_{\beta \rm rf}$.
 Belonging to the same reference frame, we may put 
 $\hat{X}^0 = \hat{X}^0_{\rm rf} = x^0$.
 The relation between
 $\hat{{\bf R}}$ and $\hat{{\bf R}}_{\rm rf}$ is achieved by a Lorentz
 transformation from the rest frame ($X^{\mu}_{\rm rf}$) to the
 moving frame where, for centre-of-mass coordinates,
 the hat applies ($\hat{X}^{\mu}_{\rm rf}$). One finds,

\medskip
 
\begin{equation} 
\hat{{\bf R}} ~ =~ \hat{{\bf R}}_{\rm rf}~  -~ \frac{{\bf S}_{\rm rf}}{Mc}
\times \frac{{\bf P}}{P^0}\, ,
\end{equation}

\bigskip

where ${\bf S}_{\rm rf}$ denotes the spin of the object in the
rest frame,  $S^{ij}_{\rm rf} = \epsilon^{ijk0}S_{k \rm rf}$, [3], [2].

Obviously, $\hat{{\bf R}}  - \hat{{\bf R}}_{\rm rf}$, i.e. the difference
vector of the centre-of-mass positions, on the one side defined in the 
reference frame where the object is moving and on the other side defined
in the rest frame but (Lorentz) transformed
to the system where the object is moving,
is orthogonal to ${\bf v}$; thus there is no Lorentz
contraction involved. Under the
assumption of positivity of energy density of the object in all
inertial frames it follows that the radius of the {\bf minimum
size} of the object, orthogonal to the spin direction, is given by [3], [4],

\medskip

\begin{equation} 
\frac{|{\bf S}_{\rm rf}|}{Mc}\, . 
\end{equation} 

\bigskip

The Eq. 8 also fits with electrons. The insertion of the spin of the electron 
results in the Compton wavelength as diameter of the area in question which is
consistent with the minimum size of positive frequency wave
functions. It is also nice to point out that even the radius of the ring singularity
of a Kerr black hole fits with Eq. 8 although in Einstein's theory of
gravity Minkowski spaces do exist only locally.

\bigskip

\section{3.~ GRAVITOMAGNETISM}

The line element of 4-dimensional curved spacetime can be decomposed into
various forms (the signature of the metric is +2),    

\medskip

\begin{align} 
ds^2 &= g_{\mu\nu} dx^{\mu}dx^{\nu} = g_{00}c^2dt^2 + 2g_{0i}cdtdx^i +
g_{ij}dx^idx^j \nonumber \\[2mm]
&= (-g_{00})[-(cdt-A_idx^i)^2 + (G_{ij} + A_iA_j)dx^idx^j] \nonumber \\[2mm]
&= - N^2c^2dt^2 + g_{ij} (dx^i+N^icdt)(dx^j+N^jcdt)\, ,
\end{align}

\bigskip

where by definition,

\medskip

\begin{align} 
A_i & \equiv \frac{-g_{0i}}{g_{00}}\, , \quad
G_{ij} \equiv \frac{-g_{ij}}{g_{00}}\, , \\[2mm]
N^i & \equiv \frac{-g^{0i}}{g^{00}}\, , \quad
N^2 \equiv \frac{-1}{g^{00}} = -g_{00} + g_{ij}N^iN^j\, ,
\end{align}

\bigskip

hold.
The functions $g^{\mu\nu}$ define the inverse metric,
$g^{\mu\lambda} g_{\lambda\nu} = \delta^{\mu}_{\nu}$.
$N$ and $N^i$ are called lapse and shift functions,
respectively.

The decomposition of the line element into the form where
$A_i$ appears is adapted to observers at rest in the given coordinate
system; for those observers equal time means: $cdt= A_idx^i$,
see Fig. 1. On the
other side, the decomposition where the lapse and shift functions appear
relates to observers at rest in ``absolute spaces'', i.e. in the spaces
defined by $t$ = const., [5]. Those observers move in the given
coordinate system according to $dx^i=-N^icdt$, see Fig. 2.

\rotatebox{90}{\resizebox{7.5cm}{!}{\includegraphics*[20cm,26.5cm]{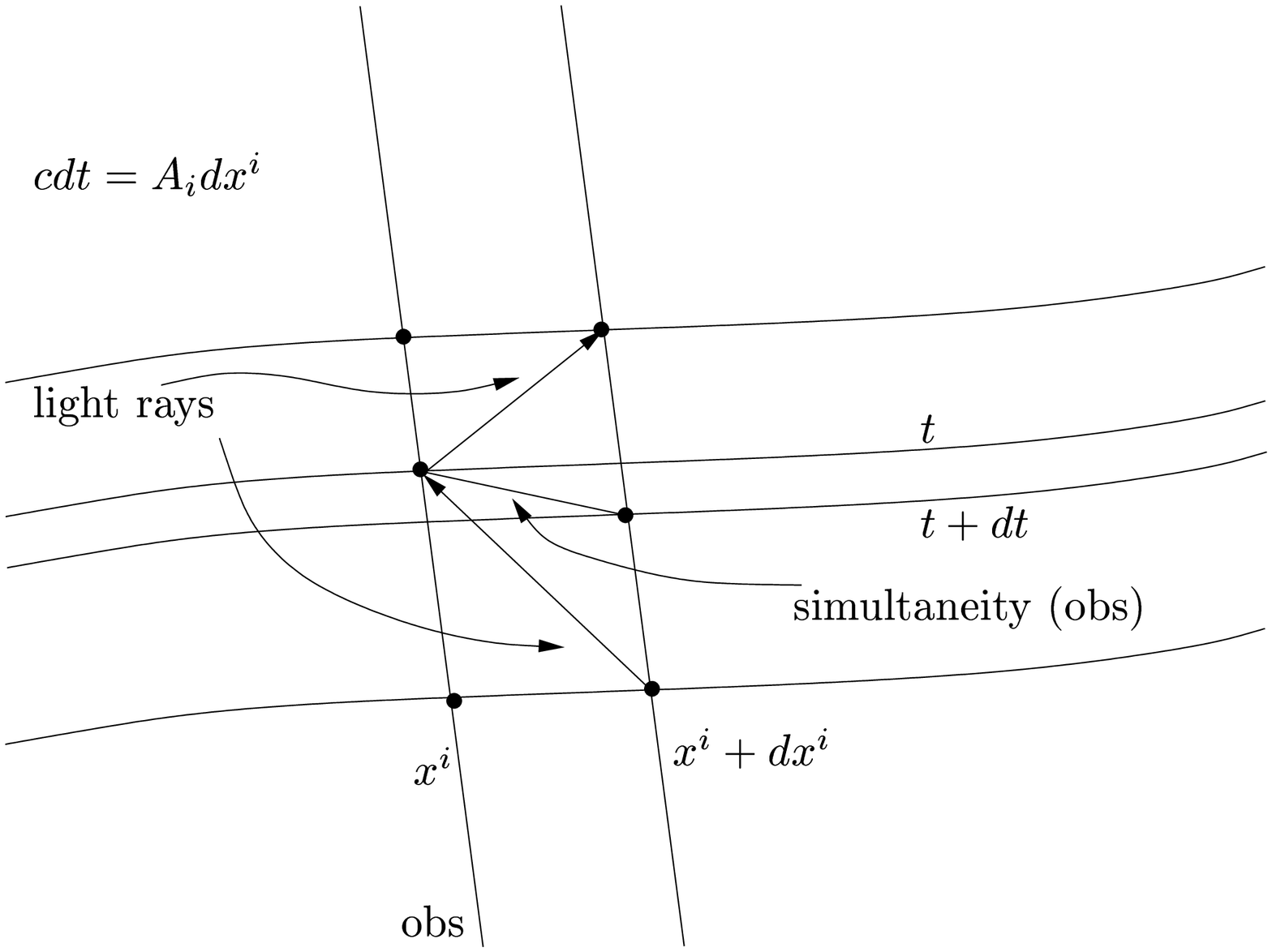}}}

Fig. 1. Shown is infinitesimal simultaneity for an observer located at
$x^i$ (Einstein synchronization).

\bigskip

The functions
$g_{0i}$, $A_i$, and $N^i$ are different representations of the 
gravitomagnetic field (this field comprises the Coriolis field as well),
respectively being denoted by ${\bf g}$, ${\bf A}$, and ${\bf N}$;
their relations are: $g_{0i}= -g_{00}A_i = g_{ik}N^k$. Obviously, 
$N^i = \gamma^{ik}g_{0k}$ is valid, whereby $\gamma^{ik}$ is the
inverse metric to $g_{ik}$, $\gamma^{il}g_{lk} =\delta^i_k$.

\rotatebox{90}{\resizebox{7.5cm}{!}{\includegraphics*[20cm,27cm]{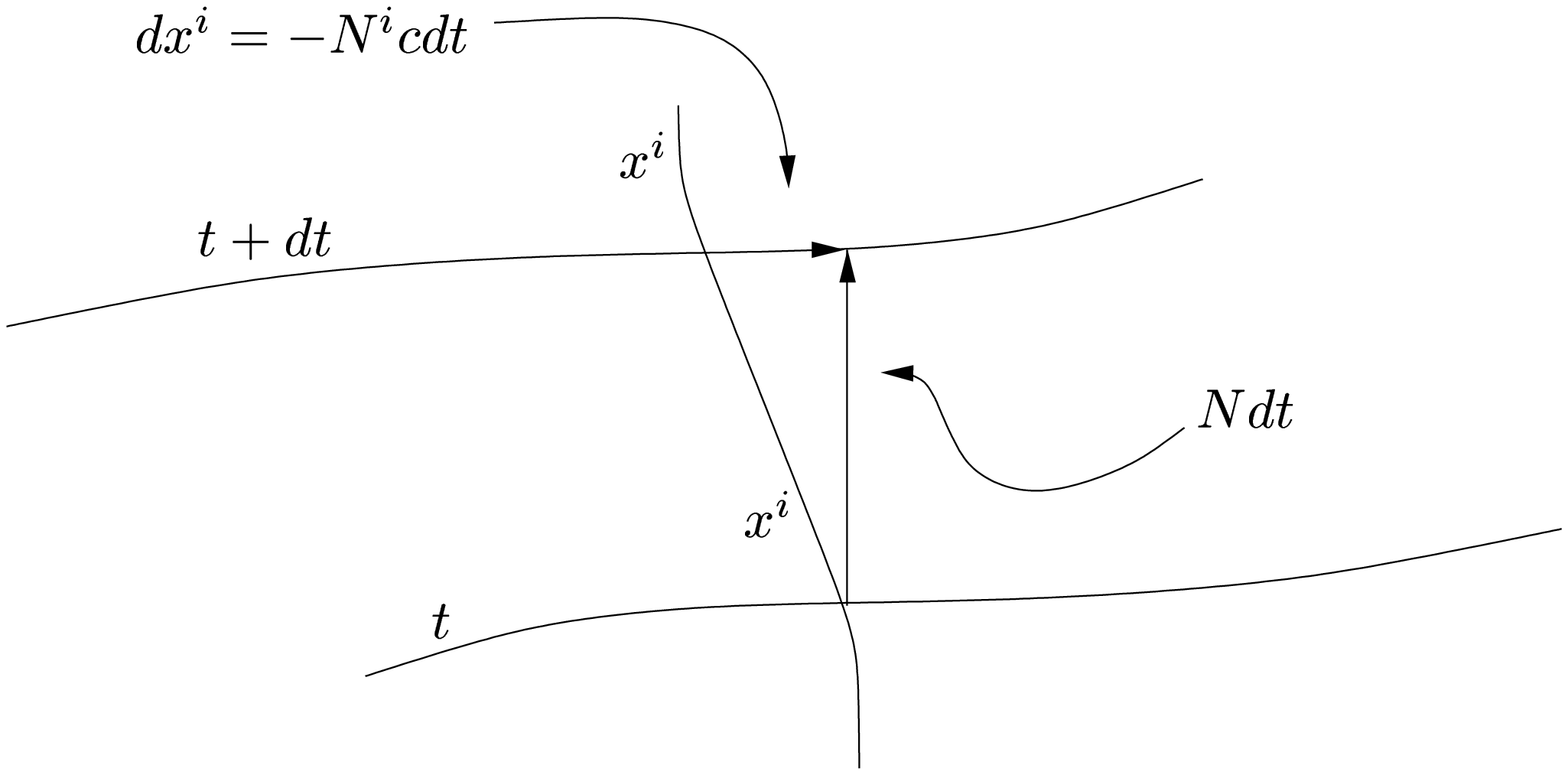}}}

Fig. 2. Shown are the lapse and shift functions. The adapted observer
moves along $Ndt$ in the spacetime.

\bigskip

In the following, typical examples of Coriolis and (genuine)
gravitomagnetic fields will be presented.

\section{3.1~ \underline{The Coriolis field}}

The line element of a rigidly rotating reference
frame, in cylindrical coordinates, takes the form,

\medskip

\begin{align}
ds^2 ~= &~ - ~(1-\frac{\Omega^2\varrho^2}{c^2})~c^2dt^2 ~+~
2\Omega\varrho^2dtd\phi \nonumber \\[2mm]
&~ + ~ (d\varrho^2 + \varrho^2 d\phi^2 + dz^2)\, .
\end{align}

\bigskip

One easily reads off, $g_{0\phi} = \Omega\varrho^2/c$, or, in
vectorial notation 

\medskip

\begin{equation}
{\bf g} ~= ~\frac{1}{c}~ {\bf \Omega} \times {\bf r}\, .
\end{equation}

\bigskip

This is the well-known Coriolis field.

In the rotating reference frame, the velocity ${\bf v} = d{\bf R}/dt$,
with  $d{\bf R} = dx^i(t)$, of a particle which is at
rest in the (global) inertial frame connected with the centre
of the rotating frame, is given by 

\medskip

\begin{equation}
{\bf v} ~=~ - ~ {\bf N}c ~=  - ~ {\bf \Omega} \times {\bf R}\, .
\end{equation}

\bigskip

It should be interesting for the reader to compare this velocity with the
corresponding one in the next section.

\section{3.2~ \underline{Dragging of inertial frames}}

The line element of the exterior field of a rotating body, in non-rotating
cylindrical coordinates, to linear order in $G$, where $G$ denotes the
Newtonian gravitational constant, reads, if the spin vector (proper
rotation vector) ${\bf S}$ is pointing in z-direction,  

\medskip

\begin{align}
ds^2 ~= & -~ (1-\frac{2GM}{c^2r})~c^2dt^2~ -~
\frac{4GS\varrho^2}{c^2r^3} dt d\phi \nonumber \\[2mm]
&~+ ~(1+\frac{2GM}{c^2r}) (d\varrho^2 + \varrho^2 d\phi^2 + dz^2)\, ,
\end{align}

\bigskip

where $r^2 = \varrho^2 + z^2$ and $S^2 = {\bf S}^2$ hold. In this case we find 
$g_{0\phi} = - 2GS\varrho^2/c^3r^3$ or, 

\medskip

\begin{equation}
{\bf g} ~=~ - ~ \frac{2G}{c^3r^3} ~ {\bf S} \times {\bf r}\, . 
\end{equation}

\bigskip

The velocity of a particle, as measured in the given coordinate system, 
which freely follows the action of the gravitational field along the azimuthal
direction (in radial direction the particle is kept fixed), is

\medskip

\begin{align}
{\bf v} ~= - ~ {\bf N}c & ~=  ~(1+\frac{2GM}{c^2R})^{-1}
\frac{2G}{c^2R^3} ~ {\bf S} \times {\bf R} \nonumber \\[2mm]
& ~= ~\frac{2G}{c^2R^3} ~ {\bf S} \times {\bf R} ~ + ~ ...
\end{align}

\bigskip

This is the famous frame-dragging effect for particle motion;
it results from the gravitomagnetic field only. The particles
show the dragging of the inertial frames in the same way as
small wood pieces may show the streamlines of flowing water.

For more insight into the ``frame-dragging'' field, we give the
orbital angular velocities of particles in circular motion in the
equatorial plane of a spinning object. The velocities read,
e.g. see [6],

\medskip

\begin{equation}
\left(\frac{d\phi}{dt}\right)^2 ~= ~\frac{GM}{R^3} ~\mp
~\frac{2GS}{c^2R^3} ~ \sqrt{\frac{GM}{R^3}}\, . 
\end{equation}

\bigskip

The naive expectation that the particle moving in the dragging direction
is the faster one is not correct, rather the particle on the retrograde
orbit is the faster (lower sign in Eq. 18). This is understandable from
the fact that the retrograde orbit needs stronger centrifugal force,
and thus higher velocity, to exist.   

\bigskip

\section{4.~ SAGNAC EFFECTS}

The differential of the phase $\psi$ of a light ray is given by $d\psi =
k_{\mu}dx^{\mu}$, where $k_{\mu} = (- \omega/c, {\bf k})$ is the
wave 4-vector with frequency $\omega$. By the aid of 
the dispersion relation $k_{\mu} k^{\mu}=0$ one can derive the integral
representation

\medskip

\begin{align}
& \psi (t,x^i;t_0,x^i_0) = \nonumber \\[2mm]
&- \int_{t_0}^t \omega dt + \int_{x_0}^x \frac{\omega}{c} (\sqrt{(G_{ij}
+ A_iA_j) dx^idx^j} + {\bf A} \cdot d{\bf r})\, ,
\end{align}

\bigskip

where ${\bf A} \cdot d{\bf r} \equiv A_idx^i$.
Along the light ray, where $ds^2=0$ holds, the phase is constant,
$\psi =0$.  Hereof,

\medskip

\begin{equation}
\int_{t_0}^t \omega dt = \int_{x_0}^x \frac{\omega}{c}
(\sqrt{(G_{ij} + A_iA_j)dx^idx^j} + {\bf A} \cdot d{\bf r})
\end{equation}

\bigskip

follows.
For time independent metric coefficients, the frequency $\omega$ is
a constant. In this case, the light path results from the simple condition that
the right side of Eq. 20 is an extremum in 3-space.
The well-known Sagnac effect in optics comes from the last term on the right
side of the Eq. 20 applied to two light rays which originate from a
specific event and end in another one. 

\section{4.1~ \underline{The Sagnac effect for atoms}}

For the proper time $\tau$ of an object we have $c^2d \tau^2 = - ds^2$.
The related action $W$ takes the form,

\medskip

\begin{equation}
W ~= ~-Mc^2 \tau\, ,
\end{equation}

\bigskip

and the quantum phase of the object is given by $W/\hbar$.
For the growth of proper time between the coordinate times
$t_0$ and $t$, one finds, 

\medskip

\begin{equation}
\tau (t;t_0) =  \int_{t_0}^t \sqrt{N^2 - g_{ij} (N^i +
\frac{v^i}{c})(N^j + \frac{v^j}{c})}~dt\, , 
\end{equation}

\bigskip

where $v^i = dx^i(t)/dt$.
The additive structure of  $cN^i + v^i$ nicely reveals the
motion of the reference frame with respect to a particle's
dragged motion in azimuthal direction, see section 3.2. 

The straight comparison with the change of phase in the light
propagation case, Eq. 20, is best achieved through the following
approximate expression, 

\medskip

\begin{align}
W = & - Mc^2 \int_{t_0}^t \sqrt{-g_{00}}(1 - (G_{ij} + A_iA_j)
\frac{v^iv^j}{2c^2} + ...) dt \nonumber \\[2mm] 
& + M c \int_ {x(t_0)}^{x(t)}  \sqrt{-g_{00}}~ {\bf A} \cdot d{\bf R}\, ,
\end{align}

\bigskip

where the phase of the  particle (atom, etc.) is given by $W/\hbar$.
The Sagnac effect for atoms results from Eq. 23 in the same way as the
Sagnac effect for light rays results from Eq. 20. By the aid
of Eqs. 21 and 23 the {\bf clock effects} originating from the gravitomagnetism can
be deduced in a straightforward manner.

\section{4.2~ \underline{Analogy with the electrodynamics}}

The gravitational Hamiltonian of a point mass reads,

\medskip

\begin{align}
H &= Nc ~ \sqrt{M^2c^2 + \gamma^{ij}P_iP_j} - c{\bf N}\cdot {\bf P}\nonumber \\[2mm] 
&= Mc^2 + \frac{1}{2M} ({\bf P} - M {\bf N}c)^2 + M \Phi ~ + ~ ...\, , 
\end{align}

\bigskip

where ${\bf P} = (P_i)$ and where
the Newtonian gravitational potential $\Phi$ has been introduced,
$g_{00} = - 1 - 2\Phi/c^2$ + ... (expansion in powers of $1/c^2$). 

The analogous expressions in the electrodynamics are,

\medskip

\begin{align}
H &= c ~ \sqrt{M^2c^2 + ({\bf P} - \frac{e}{c} {\bf A}^{\rm em})^2} + e
\Phi^{\rm em} \nonumber
\\[2mm] 
&= Mc^2 + \frac{1}{2M} ({\bf P} - \frac{e}{c}{\bf A}^{\rm em})^2 + e
\Phi^{\rm em} ~ + ~ ... 
\end{align}

\bigskip

Obviously, the analogy, valid in the weak-field slow-motion limit,
takes the form:  $e{\bf A}^{\rm em} \longleftrightarrow M {\bf N}c^2$,
$e\Phi^{\rm em}  \longleftrightarrow M\Phi$. Remember, however, the difference
in the field-generation relations: $4M{\bf v} \rightarrow -~{\bf N}c^3$,
$M \rightarrow - ~\Phi$, in the harmonic gauge, whereas 
$e{\bf v} \rightarrow {\bf A}^{\rm em}c$, $e \rightarrow \Phi^{\rm em}$,
correspondingly in the Lorentz gauge.

\section{5.~ SPIN DYNAMICS IN GRAVITY}

In the framework of the Einstein theory of gravity
the gravitational interaction Hamiltonian of two spinning objects,
linear in $G$ and linear in each spin (in this approximation,
the proper rotation of an extended object can be treated as rigid),
is obtained by the substitution, in the momentum and Hamiltonian
constraints of the Einstein field equations, of the momentum density
${\bf P}_a \delta ({\bf x} - {\bf x}_a)$, with $a=1,2$, through  
$({\bf P}_a + \frac{1}{2} {\bf S}_a \times \pmb{\nabla}_a) \delta ({\bf
x} - {\bf x}_a)$. Hereof, in the rest frame (vanishing total linear
momentum), the spin-orbit
interaction results in the form, also e.g. see [2], 

\medskip

\begin{equation}
H_{\rm SO} ~=~ \frac{2G}{c^2R^3} ~ ({\bf S} \cdot {\bf L})
~+~ \frac{3GM_1M_2}{2c^2R^3} ~ ({\bf b} \cdot {\bf L})\, , 
\end{equation}

\bigskip

where

\medskip

\begin{align}
{\bf S} & ~\equiv~ {\bf S}_1 + {\bf S}_2\, , \\[2mm]
{\bf L} & ~\equiv~ {\bf R} \times {\bf P}\, , \\[2mm]
{\bf b} & ~\equiv~ \frac{{\bf S}_1}{M_1^2} ~+~ \frac{{\bf S}_2}{M_2^2}\, , 
\end{align}

\bigskip

and where 
${\bf P} \equiv {\bf P}_1 = - {\bf P}_2$, ~${\bf R} \equiv {\bf x}_1 -
{\bf x}_2$, ~$R = |{\bf R}|$; and for the
spin-spin interaction one obtains  

\medskip

\begin{equation}
H_{\rm S_1S_2} = \frac{G}{c^2R^3} \left(\frac{3({\bf S}_1
\cdot {\bf R})({\bf S}_2 \cdot {\bf R})}{R^2} - ({\bf S}_1 \cdot {\bf
S}_2)\right)\, .
\end{equation}

\bigskip

Notice that the coordinate ${\bf R}$ is the canonical conjugate to
${\bf P}$. It relates to the particle coordinate of section 2 with 
the same name. Obviously, the total angular momentum 

\medskip

\begin{equation}
{\bf J}~ =~{\bf L} ~+~ {\bf S}\, ,
\end{equation}

\bigskip

is conserved in time,

\medskip

\begin{equation}
\frac{d{\bf J}}{dt} ~=~  \{{\bf J}, H_{\rm SO} ~+ ~H_{\rm S_1S_2}\}
~=~ 0\, ,
\end{equation}

\bigskip

where $\{,\}$ denotes the standard Poisson brackets, and 
also the absolute values of the spin vectors ${\bf S}_a ~ (a=1,2)$
do not change with time, 

\medskip

\begin{equation}
\frac{d{\bf S}^2_a}{dt} ~= ~ \{{\bf S}^2_a, H_{\rm SO} ~+~ H_{\rm S_1S_2}\}
~=~ 0\, .
\end{equation}

\bigskip

The spin-spin interaction Hamiltonian, Eq. 30, is identical with the spin(1)-spin(2)
Hamiltonian of two Kerr black holes. The latter is part of the
Hamiltonian which describes the full spin-spin interaction of two Kerr
black holes to linear and quadratic orders in $G$ and ${\bf S}$,
respectively, [5]

\medskip

\begin{equation}
H_{\rm SS}^{\rm Kerr}~ =~ \frac{GM_1M_2}{2c^2R^3} \left(\frac{3({\bf a}\cdot
{\bf R})({\bf a} \cdot {\bf R})}{R^2} - ({\bf a} \cdot {\bf a})\right)\,,
\end{equation}

\bigskip

where

\medskip

\begin{equation}
{\bf a} ~\equiv ~\frac{{\bf S}_1}{M_1} ~+ ~\frac{{\bf S}_2}{M_2}\, .
\end{equation}

\bigskip

It is interesting
to note that the spin($a$)-spin($a$) interaction terms $(a=1,2)$ are of 
monopole-quadrupole type whereas the spin(1)-spin(2) interaction
is of dipole-dipole type. The latter is independent from the masses $M_a$,
see Eq. 30. The spin-spin interaction Hamiltonian, Eq. 30, as well as
the part of the spin-orbit interaction Hamiltonian, Eq. 26, which depends on
${\bf S}$ only result from the gravitomagnetic field, the other part in
Eq. 26, depending on the masses, would also be present without
gravitomagnetic field.

For completeness we give the orbit-orbit interaction Hamiltonian 
which results from the gravitomagnetic field. In the rest frame, it
reads, cf. [7],

\medskip

\begin{equation}
H^{\rm gmag}_{\rm O_1O_2} ~=~-~ \frac{G}{2c^2R}\left(7 {\bf P}^2 +
\frac{({\bf P} \cdot {\bf R})^2}{R^2}\right) \, . 
\end{equation}

Again, the expression does not depend on the masses.

\section{5.1~ \underline{The Lense-Thirring effect}}

The (proper) Lense-Thirring effect is the precession of the orbital
plane of an object moving in the gravitomagnetic field of a spinning
central object. By the aid of the orbital angular momentum vector ${\bf L}$
and the Runge-Lenz-Laplace-Lagrange vector [7] which is defined in the rest frame
of the binary system and points from the centre-of-mass position to the
periastron of the relative orbit,  

\medskip

\begin{equation}
{\bf M} ~\equiv ~{\bf P} \times {\bf L}~ - ~\frac{GM_1^2M_2^2}{M_1+M_2}
\frac{{\bf R}}{R}\, ,
\end{equation}

\bigskip

(${\bf M} \cdot {\bf L} = 0$)
the precession of the orbit takes the form    

\smallskip

\begin{align}
<\left(\frac{d{\bf L}}{dt}\right)^{\rm S}_{\rm O}>_t & \equiv
<\{{\bf L}, H_{\rm SO}\}>_t = {\bf \Omega}_{\rm SO} \times {\bf L}\, , \\[2mm]
<\left(\frac{d{\bf M}}{dt}\right)^{\rm S}_{\rm O}>_t & \equiv
<\{{\bf M}, H_{\rm SO}\}>_t = {\bf \Omega}_{\rm SO} \times {\bf M}\, ,
\end{align}

\bigskip

where $< >_t$ denotes orbital averaging and where
the precessional frequency vector is given by 

\medskip

\begin{equation}
{\bf \Omega}_{\rm SO} ~= ~\frac{2G}{c^2} <\frac{1}{R^3}>_t \left({\bf S}_{\rm eff} - 3
\frac{({\bf L} \cdot {\bf S}_{\rm eff}){\bf L}}{L^2}\right)\, ,
\end{equation}

\medskip

with

\medskip

\begin{equation}
{\bf S_{\rm eff}} ~\equiv~ {\bf S} ~+ ~\frac{3}{4}M_1M_2 {\bf b}\, .
\end{equation}

\bigskip

If respectively $e$ and $a$ are eccentricity and semimajor axis of the
relative orbit, the averaging procedure yields, e.g. see [8], [7], 

\medskip

\begin{equation}
<\frac{1}{R^3}>_t  ~ = ~\frac{1}{a^3 (1-e^2)^{3/2}}. 
\end{equation}

\bigskip

For the LAGEOS satellite, $|{\bf \Omega}_{\rm SO}|$ results in 31 mas/yr
[9]. In this case, where the index 1 may apply to the non-spinning
satellite and the index 2 to the Earth, ${\bf S_{\rm eff}}$ is simply
given by ${\bf S}_2$.

\section{5.2~ \underline{The Schiff effect}}

The Schiff effect (also called Lense-Thirring effect for spin, or
frame-dragging effect) is the precession of a spin in the
gravitomagnetic field of a spinning central object. It is given by  

\medskip

\begin{equation}
\left(\frac{d{\bf S}_1}{dt}\right)^{\rm S}_{\rm S} ~\equiv ~ \{{\bf S}_1, H_{\rm S_1S_2}\}
~= ~{\bf \Omega}_{\rm S_2} \times {\bf S}_1\, ,
\end{equation}

\bigskip

where the precessional vector reads

\medskip

\begin{equation}
{\bf \Omega}_{\rm S_2} ~=~ \frac{G}{c^2R^3}
\left(3 \frac{({\bf R} \cdot {\bf S}_2){\bf R}}{R^2} - {\bf
S}_2\right)\, .
\end{equation}

\bigskip

For the gyroscopes of the GP-B mission an orbital-averaged spin precession 
of 41 mas/yr is predicted [10] (usually quoted as 42 mas/yr, e.g. see Fig. 5 in
[11]) to be measured with an accuracy of 0.3\%.

\section{5.3~ \underline{Gravitomagnetic field lines}}

In the following discussion about gravitomagnetic field lines
the object 2 with mass $M_2$ and spin ${\bf S}_2$ will be assumed to be
much heavier than the object 1. Then the object 1 can be treated as test
object in the gravitational field of object 2.
The shift function of object 2 reads 

\medskip

\begin{equation}
{\bf N} ~= ~\frac{2G}{c^3r^3} ~ {\bf r} \times {\bf S}_2\, . 
\end{equation}

\bigskip

Hereof the gravitomagnetic field strength ${\bf H}$, analogously to the
electrodynamics, follows in the form,  

\medskip

\begin{equation}
{\bf H} ~=~ {\bf \nabla} \times {\bf N}c = \frac{2G}{c^2r^3}
\left({\bf S}_2  - 3 \frac{({\bf r} \cdot {\bf S}_2){\bf
r}}{r^2}\right)\, .
\end{equation}

\bigskip

Obviously, 

\medskip

\begin{equation}
{\bf \Omega}_{\rm S_2} ~= ~ - \frac{1}{2}~ {\bf H}
\end{equation}

\bigskip

holds. In the exterior regime of object 2, the gravitomagnetic field strength
allows the representation 

\medskip

\begin{equation}
{\bf H} ~=~ {\bf \nabla} \lambda  
\end{equation}

\bigskip

with 

\medskip

\begin{equation}
\lambda ~ =~ \frac{2G}{c^2r^3}~ {\bf r} \cdot {\bf S_2}\, .
\end{equation}

\bigskip

As orthogonal trajectories of the $\lambda =$ const. lines, the gravitomagnetic
field lines are easily obtained. The equation for them reads,

\medskip

\begin{equation}
\frac{r_{\rm max}-r}{r_{\rm max}} ~=~ \frac{({\bf r} \cdot {\bf
S}_2)^2}{r^2S^2_2}\, ,
\end{equation}

\bigskip

where $r_{\rm max}$ is the maximum value of $r$ for a given field line
(notice: $\pmb{\nabla} \lambda ({\bf r})\cdot \pmb{\nabla} r_{\rm max}({\bf r}) = 0$).
The gravitomagnetic field strength, in the exterior regime of 
object 2, takes the form,

\medskip

\begin{equation}
{\bf H} ~=~  \frac{2GS_2}{c^2r^3 r_{\rm max}}\frac{d{\bf r}(r(\theta),
\theta)}{d{\rm cos}\theta}\, ,
\end{equation}

\bigskip

where $\theta$ is the angle between ${\bf r}$ and ${\bf S}_2$, i.e.
${\bf r} \cdot {\bf S}_2 = rS_2 {\rm cos}\theta$ and where in
the function $r(\theta)$ the $r_{\rm max}$ has to be kept
constant.

\vspace*{1.5cm}
\resizebox{8cm}{!}{\includegraphics{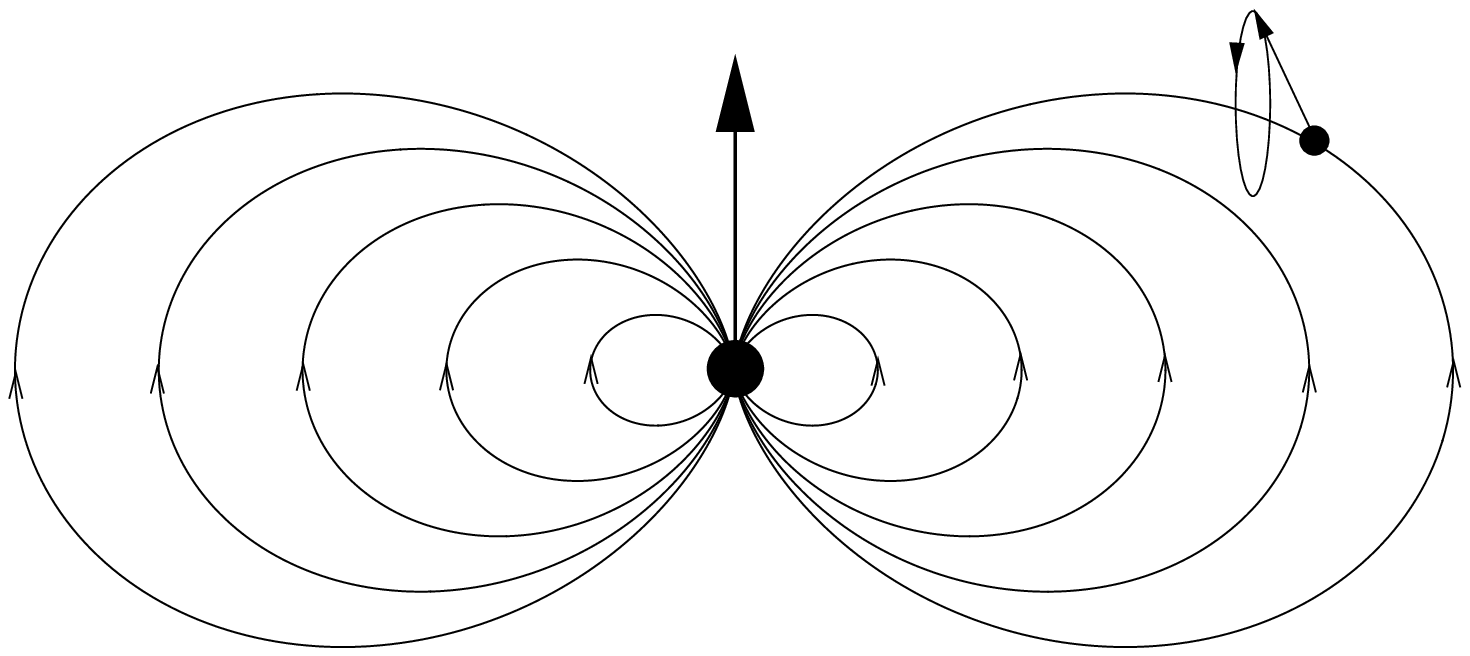}}
\setlength{\unitlength}{1cm}
\begin{picture}(0,0)(0,0)
\put(4.2,3.7){${\bf S}_2$}
\put(7.1,3.7){${\bf S}_1$}
\put(6.4,1.9){${\bf H}$}
\put(3.7,1.0){$M_2$}
\put(7.3,3.2){$M_1$}
\end{picture}

\bigskip
\bigskip

Fig. 3. Shown are the gravitomagnetic field lines of object 2 
and the spin precession of object 1 about a field line.
The object 2 is assumed to be at rest in the given coordinate system,
i.e. $M_2 \gg M_1$.

\bigskip
\bigskip

It should be remarked that the gravitomagnetic field lines in Fig. 3
differ in shape from those in other references, [5], [6], [9], [10], [11], where [5] is
closest.

\section{5.4~ \underline{The de Sitter effect}}

The de Sitter effect (also called Fokker effect, or geodetic precession)
results from 

\medskip

\begin{equation}
\left(\frac{d{\bf S}_1}{dt}\right)^{\rm S}_{\rm O}~ \equiv~  \{{\bf S}_1, H_{\rm SO}\}
~= ~{\bf \Omega}_{\rm SO}^{\rm s} \times {\bf S}_1\, ,
\end{equation}

\bigskip

where

\medskip

\begin{equation}
{\bf \Omega}_{\rm SO}^{\rm s} ~=~ \frac{2G}{c^2R^3}\left(1 +
\frac{3M_2}{4M_1}\right){\bf L}\, .
\end{equation}

\bigskip

The first term on the right side of the Eq. 53 stems from the
gravitomagnetic field.  For the Earth-Moon gyroscope, i.e. the
Earth-Moon binary system is regarded as spinning object, it holds,
19 mas/yr, e.g. see [9], and for the GP-B mission,
6,600 mas/yr are expected, e.g. see [10]. In both cases $M_1 \ll M_2$
applies (mass of Earth-Moon system vs. mass of Sun, resp. mass of satellite vs. mass of
Earth), so that only the second term on the right side of the Eq. 53 contributes.

\bigskip

{\bf Acknowledgements:} I thank G. Faye for help with the Figures
and C. L\"ammerzahl for useful discussions. 

\bigskip

\section{6.~ REFERENCES}

1. Schwinger, J., {\it Particles, Sources, and Fields I}, Perseus Books,
Reading, Massachusetts, 1998.

\bigskip

2. Barker, B.~M. and O'Connell, R.~F., The Gravitational Interaction: Spin,
Rotation, and Quantum Effects -- A Review, {\it General Relativity and
Gravitation}, Vol. 11, 149 - 175, 1979.

\bigskip

3. M{\o}ller, C., {\it The Theory of Relativity}, Oxford University Press,
Oxford 1969.

\bigskip

4. Misner, C.~W., Thorne, K.~S., and Wheeler, J.~A., {\it Gravitation}, Freeman,
San Francisco 1973.

\bigskip

5. Thorne, K.~S., Price, R.~H., and Macdonald, D.~A., (Eds.), {\it Black Holes: The
Membrane Paradigm}, Yale Univerity Press, New Haven, 1986.

\bigskip

6. Rindler, W., {\it Relativity}, Oxford University Press, Oxford, 2001.

\bigskip

7. Damour, T. and Sch\"afer, G., Higher-Order Relativistic Periastron Advances
and Binary Pulsars, {\it Nuovo Cimento B}, Vol. 101, 127 - 176, 1988.   

\bigskip

8. Landau, L.~D. und Lifschitz, E.~M., {\it Klassische Feldtheorie},
Akademie-Verlag, Berlin, 1981.   

\bigskip

9. Ciufolini, I. and Wheeler, J.~A., {\it Gravitation and Inertia},
Princeton University Press, Princeton, 1995. 

\bigskip

10. L\"ammerzahl, C. and Neugebauer, G., The Lense-Thirring Effect: From the
Basic Notions to the Observed Effects, in: {\it Gyros,
Clocks, Interferometers ...: Testing Relativistic Gravity in Space},
L\"ammerzahl, C., Everitt, C.~W.~F., and Hehl, F.~W., (Eds.),
Springer, Berlin, 2001.

\bigskip

11. L\"ammerzahl, C. and Dittus, H., Fundamental Physics in Space: A
Guide to Present Projects, {\it Ann. Phys. (Leipzig)}, Vol. 9, 1 - 57, 2000. 

\end{multicols}

\end{document}